\documentclass[superscriptaddress,showpacs,amssymb,10pt,reprint,aps,prd,longbibliography,nofootinbib,floatfix]{revtex4-1}

\usepackage{graphicx,epsfig,amssymb}
\usepackage{amsmath,amsfonts, times}
\usepackage{bm}

\usepackage{epstopdf}
\usepackage[linktocpage,colorlinks]{hyperref}
\usepackage[caption=false]{subfig}
\usepackage[usenames]{color}
\usepackage{natbib}
\usepackage{soul}
\usepackage[utf8x]{inputenc}
\usepackage[usenames]{color}
\usepackage{booktabs}
\usepackage{threeparttable}

\begin{document}
\newcommand{\beq}{\begin{equation}}
\newcommand{\eeq}{\end{equation}}
 \newcommand{\bqn}{\begin{eqnarray}}
 \newcommand{\eqn}{\end{eqnarray}}
 \newcommand{\nb}{\nonumber}
 \newcommand{\lb}{\label}
\newcommand{\PRL}{Phys. Rev. Lett.}
\newcommand{\PL}{Phys. Lett.}
\newcommand{\PR}{Phys. Rev.}
\newcommand{\PRD}{Phys. Rev. D.}
\newcommand{\CQG}{Class. Quantum Grav.}
\newcommand{\JCAP}{J. Cosmol. Astropart. Phys.}
\newcommand{\JHEP}{J. High. Energy. Phys.}
\title{Gravity-induced entanglement between two massive microscopic particles in curved spacetime: II. Friedmann- Lema\^itre-Robertson-Walker universe}

	\author{Chi Zhang}
		\email{zhangchi3244@gmail.com;}
	\affiliation{Department of Physics, Nanchang University, Nanchang, 330031, China}
	\affiliation{Center for Relativistic Astrophysics and High Energy Physics, Nanchang University, Nanchang,
330031, China}
		\author{Fu-Wen Shu}
	\email{shufuwen@ncu.edu.cn; Corresponding author}
	\affiliation{Department of Physics, Nanchang University, Nanchang, 330031, China}
	\affiliation{Center for Relativistic Astrophysics and High Energy Physics, Nanchang University, Nanchang,
330031, China}
\affiliation{GCAP-CASPER, Physics Department, Baylor University, Waco, Texas 76798-7316, USA}
\affiliation{Center for Gravitation and Cosmology, Yangzhou University, Yangzhou, China}

\begin{abstract}
In our previous work [C. Zhang and F. W. Shu, arXiv:2308.16526], we have explored quantum gravity induced entanglement of masses (QGEM) in curved spacetime, observing entanglement formation between particles moving along geodesics in a Schwarzschild spacetime background. We find that long interaction time induces entanglement, even for particles with microscopic mass, addressing decoherence concerns. In this work, we build upon our previous work  [C. Zhang and F. W. Shu, arXiv:2308.16526] by extending our investigation to a time-dependent spacetime. Specifically, we explore the entanglement induced by the mutual gravitation of massive particles in the Friedmann-Lema\^itre-Robertson-Walker (FLRW) universe. Through calculations of the phase variation and the QGEM spectrum, our proposed scheme offers a potential method for observing the formation of entanglement caused by the quantum gravity of massive particles as they propagate in the FLRW universe. Consequently, our research provides fresh insights into the field of entanglement in cosmology.
\end{abstract}
	
\date{\today}
	
\maketitle
		
\section{Introduction}
There has been a long-standing debate about whether the gravitational field is a quantum field, and to date we still have no effective way to give a clear answer. In the past few years, we have gone through a boom of interest to use desk-top experiments to explore quantum gravity effects. Many of these experiment proposals date back to a thought experiment of Feynman \cite{feynman} that if one massive particle is put in a superposition of two locations by a Stern-Gerlach device, then gravitational field would also form a superposition state. Based on this idea, Bose and Marletto et al \cite{r2,r3} proposed a plan to use quantum entanglement to detect quantum gravity effects, which is known as quantum gravity induced entanglement of masses (QGEM). In the QGEM device, if these two particles in the spatial superposition state are entangled under the action of their gravitational fields, then it can be inferred that the gravitational field has quantum properties based on the principle of quantum information (see some recent progress \cite{Christodoulou:2018cmk,Capolupo:2019peg,Belenchia:2018szb,Carney:2018ofe,Marshman:2019sne,Westphal:2020okx,Carlesso:2019cuh,Danielson:2021egj,Christodoulou:2022mkf,Cho:2021gvg,Matsumura:2020law,Miki:2020hvg,Howl:2023xtf,Yant:2023smr,Feng:2023krm,Schut:2023eux,Fragolino:2023agd,Li:2022yiy,Bose:2022uxe,He:2023hys}). However, due to the harsh experimental parameters and quantum states are highly susceptible to decoherence, these experimental designs are difficult to implement in practice. 

Very recently, we developed an astronomical version of QGEM in curved spacetime, focusing on the observation of entanglement formation between pairs of particles moving along geodesics in a Schwarzschild spacetime background \cite{Zhang:2023llk}. We find that the long interaction time between particles can induce entanglement even for particles with microscopic mass, providing a possible solution of the issue of the decoherence. In this work \cite{Zhang:2023llk}, we also found that pairs of superposed particles exhibit varying degrees of entanglement depending on different motion parameters. Specifically, entanglement formation is more likely to occur when the proper motion time is longer, the geodesic deviation rate is appropriate, and the mass of the galaxy is smaller.

In the previous work, we considered the QGEM in the Schwarzschild spacetime, which is static. In this work, we would like to investigate how QGEM evolves in a time-dependent spacetime. To be more specific, we will investigate entanglement induced by mutual gravitation between massive particles in the FLRW universe. 

There have been some previous studies on entanglement in cosmology. For instance, the production of entanglement in quantum fields due to the expansion of the underlying spacetime was shown in \cite{Martin-Martinez:2012chf}. The entanglement entropy of cosmological perturbations in the very early universe has also been investigated \cite{Brahma:2020zpk}. In the realm of photon entanglement, various avenues to utilize entanglement in order to investigate the nature of the universe have been explored. Such as schemes using entangled photons to study the origins of cosmic microwave background (CMB) fluctuations \cite{Kiefer:2008ku,Martin:2007bw,Martin:2015qta,Martin:2016tbd} and investigations involving entangled photons produced from the decay of cosmic particles to test quantum mechanics \cite{Lello:2013bva,Valentini:2006yj}. In addition, astronomers have even attempted to detect entangled photons that have traveled different paths through the cosmos and reached Earth \cite{Chen:2017cgw}.

However, investigations into the entanglement induced by the gravitational interaction of massive particles are still lacking. Our current knowledge does not include an understanding of how QGEM evolves in the FLRW universe. Filling this gap is one of our main motivations for conducting this research. Intuitively, pairs of adjacent particles propagating in the FLRW universe are expected to become entangled due to their gravitational field. However, determining the specific factors influencing their entanglement requires detailed calculations. The entanglement induced by quantum gravity, which we will present in this article, offers a fresh perspective for studying entanglement in cosmology.

This paper is organized as follows. In Sec. II, we present the overall method for quantifying the entanglement formation induced by the QGEM mechanism in the context of FLRW cosmology, and also study the influencing factors of the entanglement phase. In Sec. III, incorporating more realistic observational scenarios, we present the characteristic spectral lines of the entanglement witness as varying functions of the initial motion parameters of the particle pair and the observed kinetic energy. In Sec. IV, conclusions and discussions are made.

Throughout this paper, we will adopt the natural units system, $c=G=1$, in order to simplify the calculations. Except that the physical quantities with units are given in the SI system of units.

\section{Entanglement generation in FLRW spacetime}

For a flat FLRW universe, the metric is
\beq\label{19}d{s^2} = d{t^2} - {a(t)^2}(d{r^2} + {r^2}d{\theta ^2} + {r^2}{\sin ^2}\theta {d^2}\varphi ),\eeq
where $a(t)$ is the scale factor of the FLRW universe and it obeys
\beq\label{22}a'{\left( t \right)^2} = \frac{8}{3}\pi a{\left( t \right)^2}\rho \left( t \right).\eeq
The general density in \eqref{22} is given by 
\beq\label{21}\rho \left( t \right) = \frac{{a_0^3{\rho _{{M_0}}}}}{{a{{\left( t \right)}^3}}} + \frac{{a_0^4{\rho _{{R_0}}}}}{{a{{\left( t \right)}^4}}} + {\rho _{{v_0}}},\eeq
where
\beq\label{20}{\rho _{{v_0}}} = \frac{{3H_0^2{\Omega _\Lambda }}}{{8\pi }},{\rho _{{M_0}}} = \frac{{3H_0^2{\Omega _M}}}{{8\pi }},{\rho _{{R_0}}} = \frac{{3H_0^2{\Omega _R}}}{{8\pi }}.\eeq
Here $\Omega_{\Lambda}$, $\Omega_M$ and $\Omega_R$ are the density parameters for dark energy, matter and radiation, respectively. $H_0$ is the Hubble constant.

It was shown in \cite{r2,r3} that nearby neutral massive particle pairs exhibit phase growth under their self-interaction of gravity. The phase growth in general is given by \cite{Zhang:2023llk}
\beq\label{dphi}
\delta \phi  =  - \frac{{{m_0}\delta \tau }}{\hbar } = \frac{{{m_0}^2}}{\hbar }\int {\frac{1}{{d\left( \tau  \right)}}d\tau },
\eeq
where $m_0$ and $\tau$ are, respectively, the static mass and the proper time of the particles. $d\left( \tau  \right)$ represents the distance between two particles and is mainly determined by the geodesic deviation. Eq. \eqref{dphi} shows that the key to calculate phase change is to obtain the geodesic deviation distance $d\left( \tau  \right)$ of the particle pair.

To do this, we use the tetrad formalism and introduce the following orthogonal normalized tetrad of flat FLRW universe
\begin{equation}\label{tetrad}
\begin{aligned}
&{e_{(0)}}^{\mu} = (\partial t)^{\mu},\ \  \ \ \ \ \ \  \ \ \ \ \
{e_{(1)}}^{\mu} = \frac{1}{{a\left( t \right)}}(\partial r)^{\mu},\\
&{e_{(2)}}^{\mu} = \frac{1}{{a\left( t \right)r}}(\partial \theta)^{\mu}, \ \
{e_{(3)}}^{\mu} = \frac{1}{{a\left( t \right)r\sin \theta }}(\partial \varphi)^{\mu},
\end{aligned}
\end{equation}
so that they form an orthogonal base
\beq\label{normal}
g_{\mu\nu}{e_{(a)}}^{\mu}{e_{(b)}}^{\nu}=\eta_{(a)(b)},
\eeq
where $(a=0,1,2,3)$ and $\eta_{(a)(b)}=\text{diag}.(1,-1,-1,-1)$. 

Recalling the geodesic equations in terms of four-velocity $u^{\mu}(\tau )$ is
\begin{equation}\label{geo1}
{u^{\mu}} \circ ({e_{(a)} }_{\nu}{u^{\nu}}) + ({e^{(b)}}_{\nu}{u^{\nu}}){(d{e_{(b)} })_{\rho\sigma}}{u^{\rho}}{e_{(a)} }^{\sigma} = 0.
\end{equation}
After substituting Eq. \eqref{tetrad} into Eq. \eqref{geo1}, it reduces to
\beq\label{geo2}\begin{split}
\ddot t\left( \tau  \right) + a^2(t)H(t){{\dot r(\tau)}^2} = 0,\\
\ddot r\left( \tau  \right) + 2H(t)\dot t\left( \tau  \right)\dot r\left( \tau  \right) = 0,
\end{split}\eeq
where the dot represents differentiation with respect to $\tau$. $H(t)$ in Eq. \eqref{geo2} is defined as $H\equiv \frac{da(t)/dt}{a(t)}$, which is determined by the Friedmann equation:
\beq\label{friedm}
H^2=H_0^2\left(\Omega_{\Lambda}+\Omega_{M}\left(\frac{a_0}{a}\right)^3+\Omega_{R}\left(\frac{a_0}{a}\right)^4\right),
\eeq
The full geodesics of the FLRW universe can be obtained by solving the geodesic equation \eqref{geo2} and the Friedmann equation \eqref{friedm}.

Now let us turn to calculate the geodesic deviation vectors, from which one can read off values of $d(\tau)$, a key term in calculating $\delta\phi$ as shown in Eq. \eqref{dphi}. We obtain these vectors by solving the geodesic deviation equations in tetrad formalism. Note that unlike the fixed tetrad used in Eq. \eqref{tetrad} for geodesics, the tetrad here must be parallelly transported along the geodesics \cite{r7}, namely, ${{{\tilde{e}}_{(a)}}^{\mu}}{}_{;\nu}v^{\nu}=0$, where $v^{\nu}$ is the tangent vector of the geodesics. This means that the orientations of the axes are fixed and they are non-rotating as determined by local dynamical experiments. It turns out that the following choice of tetrads are appropriate
\begin{equation}\label{tetrad2}
\begin{aligned}
&{\tilde{e}_{(0)}}^{\mu} = \dot t(\partial t)^{\mu} + \dot r(\partial r)^{\mu},\\
&{\tilde{e}_{(1)}}^{\mu}  = a\left( t \right)\dot r(\partial t)^{\mu} + \frac{{\dot t}}{{a\left( t \right)}}(\partial r)^{\mu},\\
&{\tilde{e}_{(2)}}^{\mu} = \frac{1}{{a\left( t \right)r}}(\partial \theta)^{\mu},\\
&{\tilde{e}_{(3)}}^{\mu} = \frac{1}{{a\left( t \right)r\sin \left( \theta  \right)}}(\partial \varphi)^{\mu}.
\end{aligned}
\end{equation}
Note that in order for the above tetrad satisfying the orthogonal normalized condition \eqref{normal}, one extra condition should be imposed
\beq\label{nc}
{\dot t^2} - a{\left( t \right)^2}{\dot r^2}=1.
\eeq
The geodesic deviation equation in the above frames is given by
\beq\label{14}\frac{{{d^2}{w^{(a)} }}}{{d{\tau ^2}}} + {k^{(a)} }_{(b)} {w^{(b)} } = 0, \eeq
where $w^{(a)}$ is geodesic deviation vector in tetrad form
\beq
w^{(a)}={\tilde{e}^{(a)} }_{\mu}w^{\mu},
\eeq
and
\beq\label{35}{k^{(a)} }_{(b)}  =  - {R^{\mu}}_{\nu\rho\sigma}{\tilde{e}^{(a)} }_{\mu}{v^{\nu}}{v^{\rho}}{\tilde{e}_{(b)} }^{\sigma}.
\eeq

Substituting the tetrad \eqref{tetrad2} and \eqref{nc} into above definition, one finds that the nonvanishing components of ${k^{(a)} }_{(b)}$ are
\bqn\label{25}
{k^{(1)} }_{(1)}& =&  - \frac{1}{a(t)}\frac{dH}{dt},\\
{k^{(2)} }_{(2)} &=&  - \frac{1}{a(t)}\frac{dH}{dt}{\dot t}^2 +H^2{\dot r}^2.
\eqn

To simplify our analysis, we assume that $w^{\mu}$ is orthogonal to ${e_{(0)}}^{\mu}$, giving us $w^{(0)}=0$. Since the background exhibits spherical symmetry, the second and third components of $w^{(a)}$ are simply symmetric angular components. Thus, we can choose a suitable frame where the initial value of $w^{(3)}$ is zero. Notably, Eq. \eqref{14} is a homogeneous equation. Consequently, if $\hat{w}^{(a)}$ is a solution, so is $\kappa\hat{w}^{(a)}$, where $\kappa$ is an arbitrary nonzero constant. Therefore, if we initially set $w^{(3)}$ and its first time derivative to zero, it will remain zero throughout. Ultimately, we are left with two non-zero components: $w^{(1)}$ and $w^{(2)}$.

By substituting these components into the geodesic deviation equation \eqref{14} and performing numerical calculations with specified initial conditions, one can determine the distance between the pair of particles
\beq\label{27}
d(\tau ) = \sqrt {{w^{(1)}}{{\left( \tau  \right)}^2} + {w^{(2)}}{{\left( \tau  \right)}^2}} \Delta l.
\eeq
The change in proper time is thus given by
\beq\label{28}
\delta \tau  =  - {m_0}\int_0^{\tau '} {\frac{1}{{d\left( \tau  \right)}}d\tau },
 \eeq
 which leads to the change in phase $\delta \phi  =  - \frac{{{m_0}\delta \tau }}{\hbar }$.

In our numerical simulations, the particle's static mass ${m_0}$ is also set to be ${10^{ - 25}}\text{kg}$. And we adopt the following data: $\Omega_{\Lambda}=0.6935$, $\Omega_m=0.3065$, $\Omega_R=0$ and $H_0=67.6~\text{km} \cdot \text{s}^{ - 1} \cdot \text{Mpc}^{ - 1}$ as suggested in Plank\textbf{ 2018 } data \cite{Aghanim:2018eyx}. The scale factor at present ${a_0}$ is set to 1.

One point deserves mention is that solving the geodesic deviation equations of present case needs assign both the initial temporal coordinate $t_0$ and the initial spatial coordinate $r_0$. In the following simulations, we choose four typical sets of geodesics and their initial values which are adopted in the following numerical simulations can be found in Table \ref{initial}. Note that throughout the paper, $z$ represents the redshift of the source. It is not difficult to generalize the simulations to the cases where  $w_0^{(1)}\neq w_0^{(2)}$ and $\dot{w}_0^{(1)}\neq \dot{w}_0^{(2)}$.

\begin{figure}\centering
\includegraphics[scale=0.5]{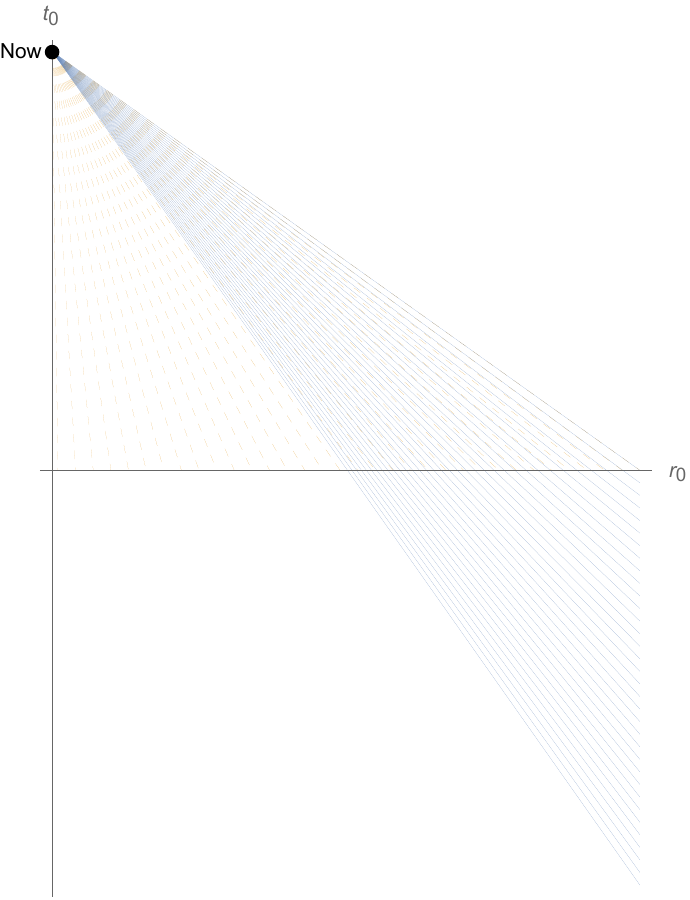}
\caption{Two different assignments of initial values for $r_0$ and $t_0$. Solid one represents that particles start from different time $t_0$ for a fixed radial coordinate $r_0$. Dashed line represents that particles start from different initial radial coordinates $r_0$ at a fixed initial time slice $t=t_0$. The black point marked "Now" represents the spacetime point where we receive particles.}\label{f7}
\end{figure}

First of all, let us focus on two assignments. One is that an initial time slice ($t=t_0$) is chosen, while particles start at different radial coordinates $r=r_0$ (that is, a snapshot of different place at time $t_0$). The other is an initial radial coordinate $r_0$ is fixed, while initial time coordinate is free (i.e., focusing on one fixed place for different time). These two series of geodesics are illustrated in Fig.\ref{f7}.

\begin{table}[h]\centering
	\begin{threeparttable}
		\caption{Sets of initial values for numerical simulations}
		\begin{tabular}{ccccc}\hline\hline
			Sets & redshift $z$& $r_0$ & $w_0^{(1)}=w_0^{(2)}$ & $\dot{w}_0^{(1)}=\dot{w}_0^{(2)}$ \\
			\hline
			S1   & 3.94  &    $4.5 \times {10^{17}}$       & $ \frac13\times 10^{-18}$  & 0 \\
			S2   & 3.94  &    $4.5 \times {10^{17}}$       & $ \frac23\times 10^{-18}$  & 0  \\
			S3   & 3.94  &    $4.5 \times {10^{17}}$       & $ \frac13\times 10^{-18}$  & $10^{-35}$  \\
			S4   & 3.94  &    $4.5 \times {10^{17}}$       & $ \frac13\times 10^{-18}$  & $-10^{-35}$ \\
			\hline\hline
		\end{tabular}\label{initial}
		\begin{tablenotes}
			\item[1] Note that this is in natural unit system. In SI system, $r_0$, $w_0^{(i)}$ and $\dot{w}_0^{(i)}$ should be multiplied by $c=3\times 10^8 \text{m/s}$. For example, for S1, we have $r_0=1.35 \times {10^{26}}\text{m} $, $w_0^{(1)}=w_0^{(2)}=10^{-10} \text{m}$ ($\gg \lambda_c\sim 10^{-15}$m, where $\lambda_c$ is the Compton wavelength of the particle), and  $\dot{w}_0^{(1)}=\dot{w}_0^{(2)}=0$.
		\end{tablenotes}
	\end{threeparttable}
\end{table}

\begin{figure}\centering
\subfloat[]{\includegraphics[scale=0.4]{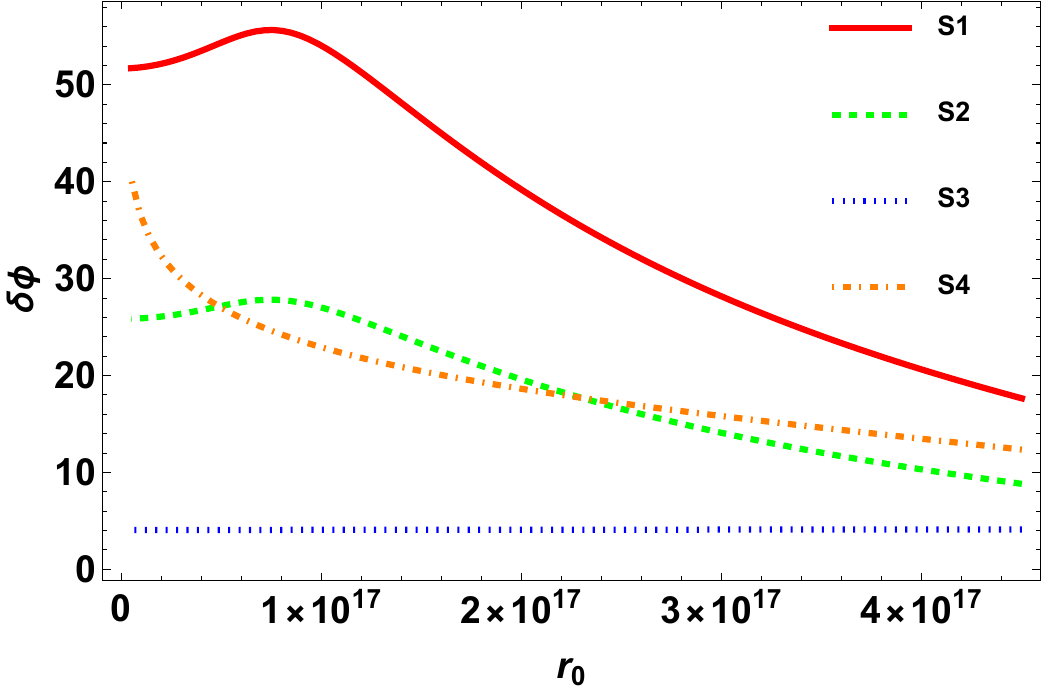}\label{f2a}}\\
\subfloat[]{\includegraphics[scale=0.4]{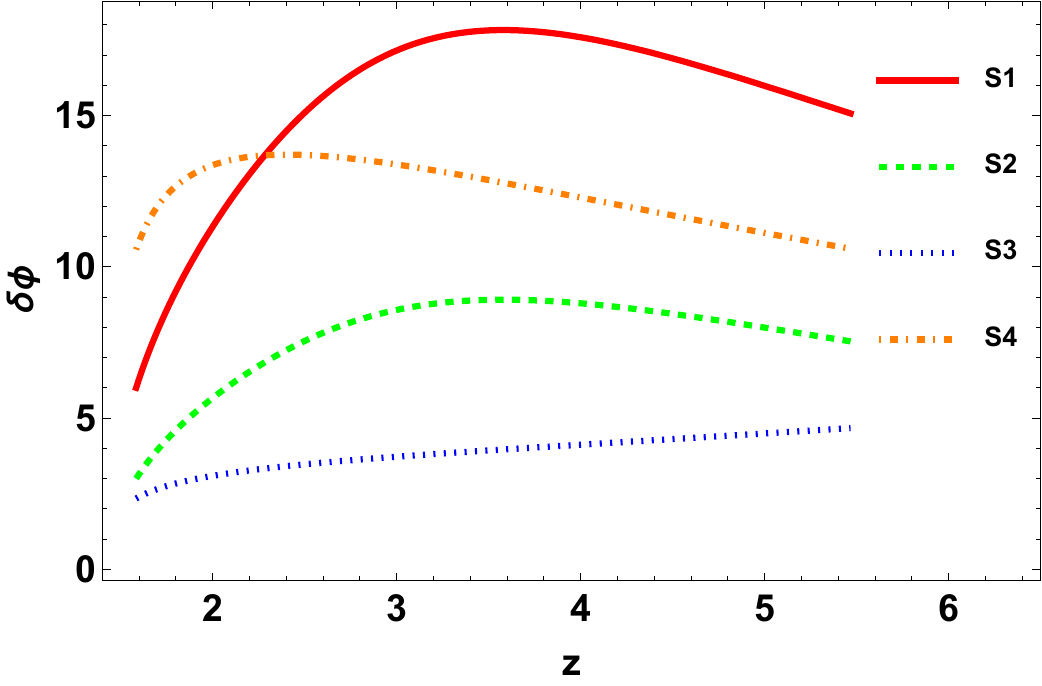}\label{f2b}}
\caption{Phase change $\delta\phi$ as a function of $t_0$ and $r_0$. (a) $\delta\phi$ against $r_0$ with ${t_0}$ fixed at \(z = 3.94\). (b) $\delta\phi$ against $z$ (or $t_0$) with ${r_0}$ fixed at ($r_0 = 4.5 \times {10^{17}}$ in natural units system).}\label{f2}
\end{figure}

Figure \ref{f2} presents a plot illustrating the phase change of two series of geodesics as a function of $t_0$ (with fixed $r_0$) or $r_0$ (with fixed $t_0$). In Fig.\ref{f2a}, we observe that for geodesics S1 and S2, the phase change initially increases to a maximum value and then gradually decreases. On the other hand, for S3 and S4, the phase change diminishes as the radial coordinate increases, accompanied by a decrease in descent speed. Fig.\ref{f2b} reveals that the phase change initially grows with increasing initial time coordinate, followed by a subsequent decrease. This behavior arises due to the interplay between increased proper time and increased separation during the movement as the initial time coordinate increases. Furthermore, we can conclude that specific initial conditions, such as a smaller radial coordinate and an appropriate time coordinate, yield the maximum phase change.

\begin{figure}\centering
\includegraphics[scale=0.4]{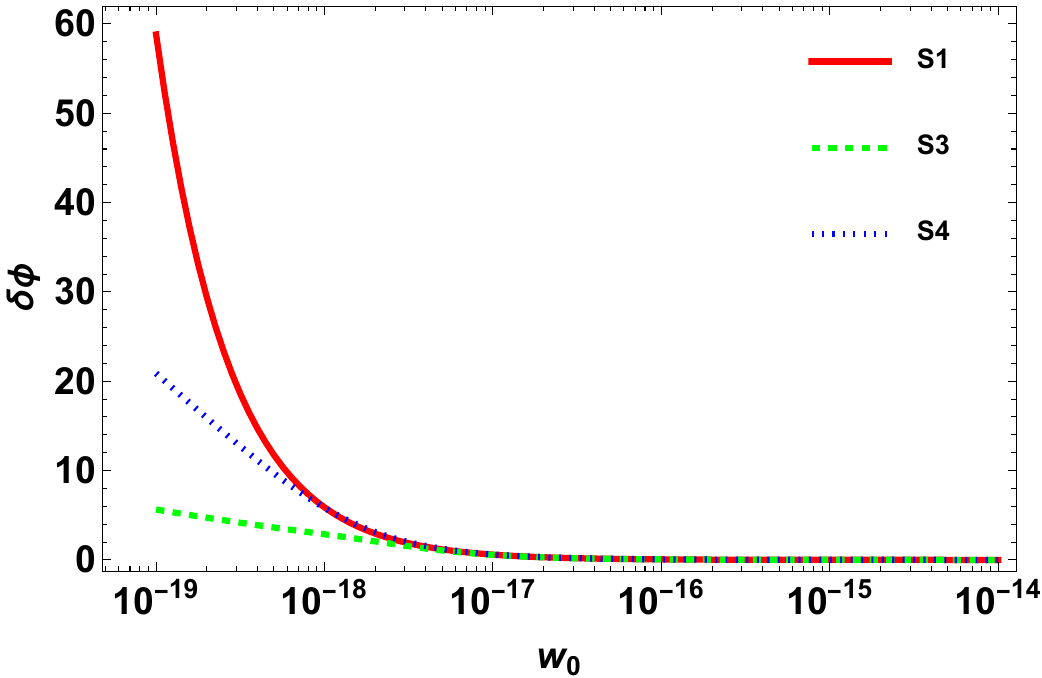}
\caption{Phase change $\delta\phi$ as a function of initial geodesic deviation $w_0$. Where we assume ${w_0^{(1)}} = {w_0^{(2)}} ={w_0}$ (in natural units system).}\label{f9}
\end{figure}

According to \cite{Christodoulou:2018cmk}, strict ranges must be set for particle mass, spacing distance, and action time. However, in our scenario, the particle's static mass can be significantly smaller, and the spacing distance can be much larger. For instance, when the initial time coordinate is $1.5 \times 10^{17}$ and the radial coordinate is $4.5 \times 10^{17}$, the phase change of particle pairs with a static mass of $1 \times 10^{-20}$ kg, separated by one meter initially, and with no initial deviation rate, can reach a substantial value of 0.361988. Such an experiment would be impractical to conduct on Earth due to the extremely light particles and the large spacing distance required. The relationship between phase change and initial geodesic offset distance under these conditions is depicted in Fig.\ref{f9}.

\begin{figure}\centering
\subfloat[]{\includegraphics[scale=0.4]{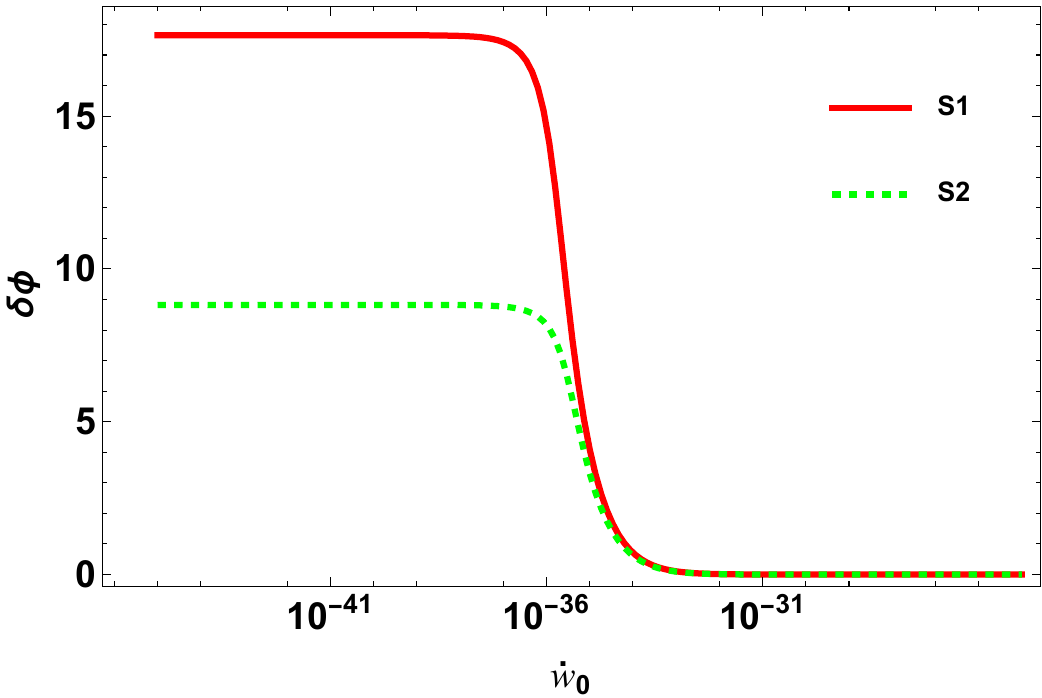}\label{f3a}}\\
\subfloat[]{\includegraphics[scale=0.4]{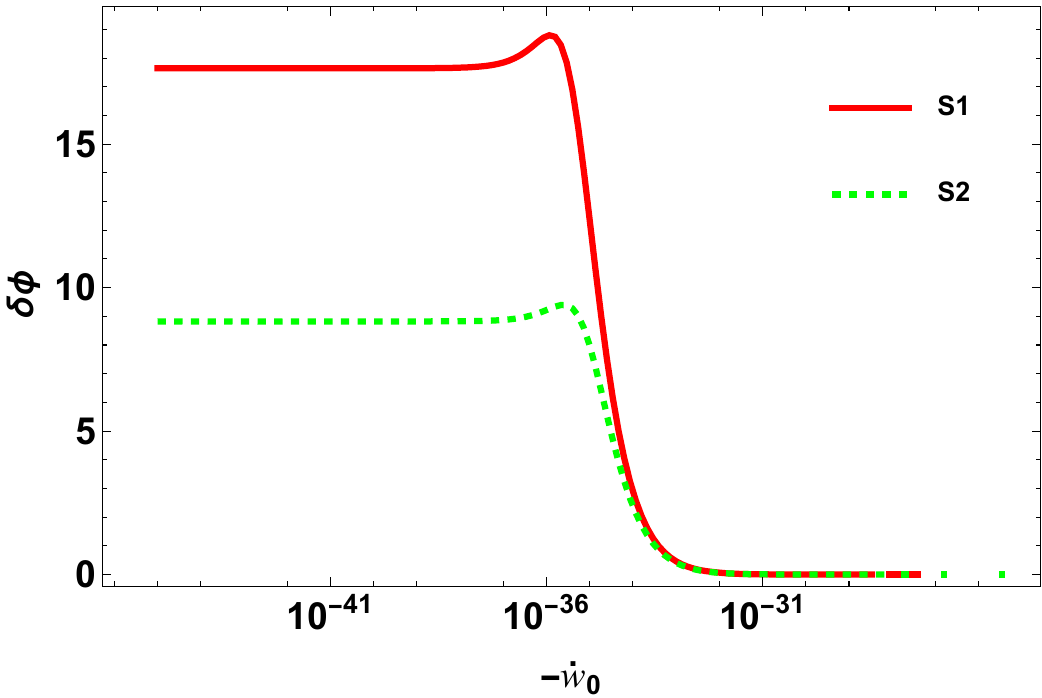}\label{f3b}}
\caption{Phase change $\delta\phi$ as a function of initial geodesic deviation velocity $\dot{w}_0$. (a) Initial geodesic deviation velocity is positive, ${\dot w_0^{(1)}} = {\dot w_0^{(2)}}> 0$. (b) Initial geodesic deviation velocity is negative, ${\dot w_0^{(1)}} = {\dot w_0^{(2)}}< 0$. In both cases, we take ${t_0} = 5 \times {10^{16}},{r_0} = 4.5 \times {10^{17}}, {w_0^{(1)}} = {w_0^{(2)}} = \frac{1}{3} \times {10^{ - 18}}$ (in natural units system).}\label{f3}
\end{figure}

We now investigate the influence of the initial velocity of the geodesic deviation vector on the phase change. By keeping the starting point of the particles fixed at a spacetime point and varying the initial velocity, we can analyze this relationship. The specific initial conditions are provided in the caption of the Fig.\ref{f3}.

In Fig.\ref{f3a}, we observe a concave phase change curve that indicates a decreasing rate of phase change as the positive initial deviation velocity increases. This behavior can be easily understood: the positive initial deviation velocity causes an increase in the spacelike distance, as depicted in Eq. \eqref{28}, leading to a smaller phase change.

On the other hand, Fig.\ref{f3b} demonstrates that the phase change initially increases with a negative initial deviation velocity and then gradually decreases until it reaches nearly zero. This can be explained by considering the following: with a relatively small negative velocity, the two particles remain close neighbors for most of the time. However, with a relatively large negative velocity, they move apart, resulting in a considerable separation between them by the end.

The entanglement phase is closely related to the cosmological model we assume, and using different cosmological models will lead to different entanglement curves. For example, the Brans–Dicke (BD) theory is a well studied scalar–tensor theory of gravity. The Lagrangian of BD theory is 
\begin{equation}
L_{\text{BD}}=\sqrt{-g}\Big[\frac{\phi^2 R}{8w}-\frac12 \nabla_{\mu}\phi\nabla^{\mu}\phi+L_M(\varphi)\Big],
\end{equation}
where $w$ is an arbitrary constant.  The present limits of the constant $w$ based on time delay experiments \cite{Gaztanaga:2000vw} require $w>10^4$. The BD theory approaches General Relativity (GR) in the limit $w\rightarrow \infty$ \cite{Clifton:2011jh}. 

The field equation for flat FLRW cosmology in BD theory is \cite{Clifton:2011jh}:
\bqn
{H^2} = \frac{{8\pi \rho }}{{3\phi }} - H\frac{{\dot \phi }}{\phi } + \frac{\omega }{6}\frac{{{{\dot \phi }^2}}}{{{\phi ^2}}},\label{bd1}\\
\frac{{\ddot \phi }}{\phi } = \frac{{8\pi }}{\phi }\frac{{(\rho  - 3P)}}{{(2\omega  + 3)}} - 3H\frac{{\dot \phi }}{\phi },\label{bd2}
\eqn
where $\phi$ is the scalar field and $H$ is the Hubble constant that evolves over time in BD theory. In deriving \eqref{bd1} and \eqref{bd2}, we have assumed that $\phi$ does not couple to the matter field $L_M$ and we are considering the classical perfect fluid in the matter field.

\begin{figure}\centering
	\subfloat[]{\includegraphics[scale=0.4]{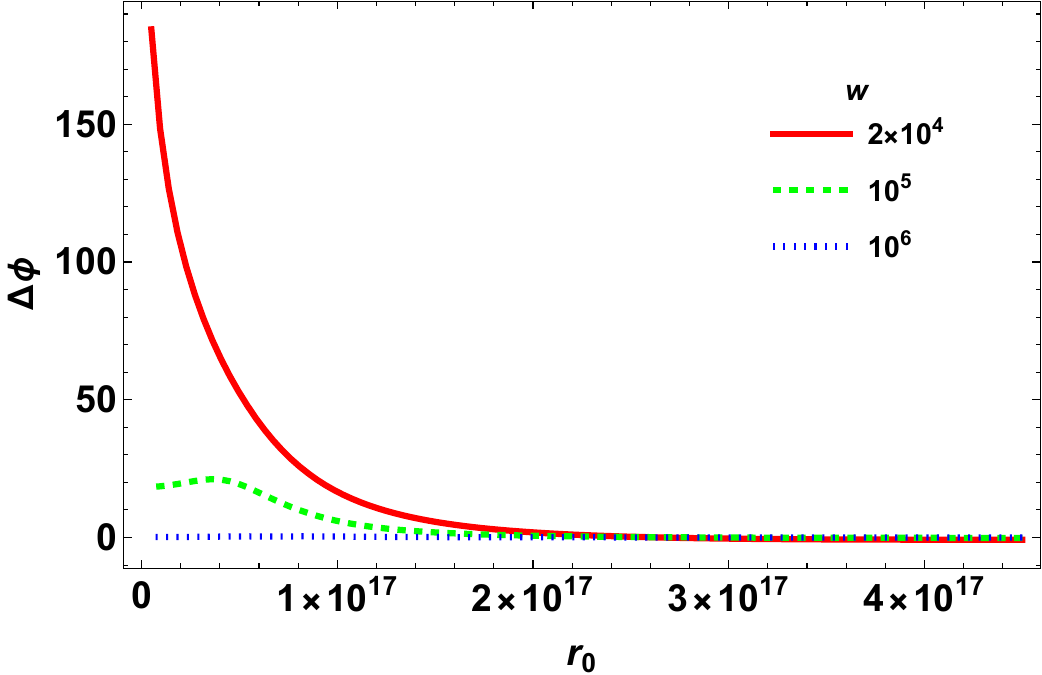}\label{f4a}}\\
	\subfloat[]{\includegraphics[scale=0.4]{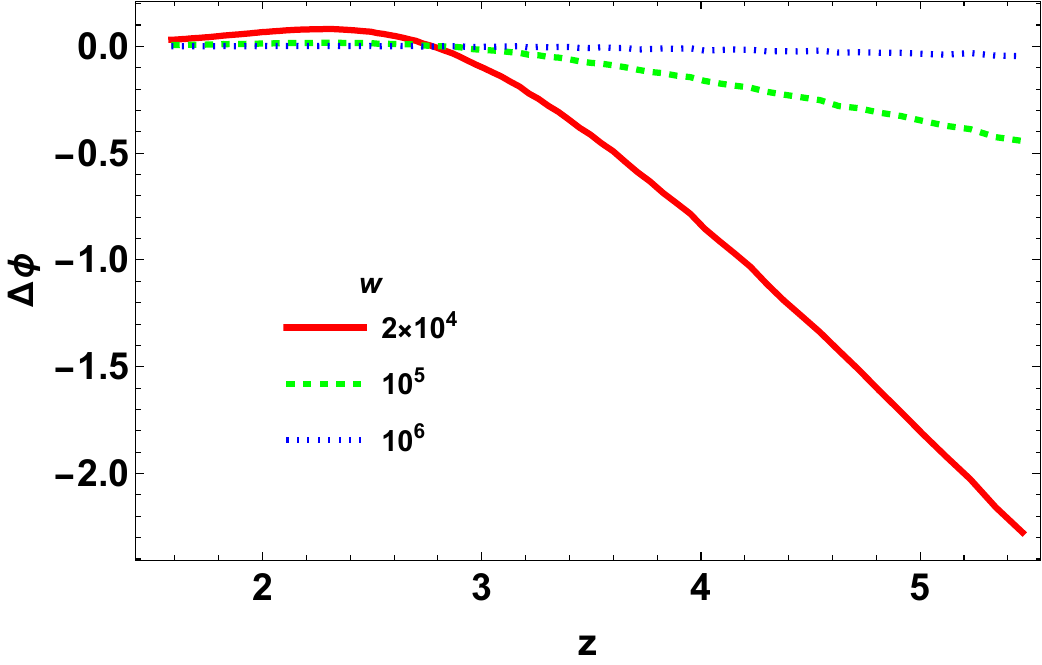}\label{f4b}}
	\caption{Phase change difference $\Delta\phi$ between BD theory and GR as a function of $t_0$ and $r_0$. The $\Delta \phi$ is defined as $\Delta \phi  = \delta {\phi _{BD}} - \delta {\phi _{GR}}$. (a) $\Delta\phi$ against $r_0$ with ${t_0}$ fixed at \(z = 3.94\). (b) $\Delta\phi$ against $z$ (or $t_0$) with ${r_0}$ fixed at ($r_0 = 4.5 \times {10^{17}}$). In both cases, we take $ {w_0^{(1)}} = {w_0^{(2)}} = \frac{1}{3} \times {10^{ - 18}}$( $\approx 10^{-10}$m), $ {{\dot w}_0^{(1)}} = {{\dot w}_0^{(2)}} = 0$ (in natural units system).}\label{f4}
\end{figure}

Following the algorithm in the previous section, we also calculated the phase shift of the two series of geodesics in Fig.\ref{f2} and compared them with those in GR. Unlike GR, here the Newton's constant is a variable: $G\left( t \right) = \frac{{2w + 4}}{{\left( {2w + 3} \right)\phi \left( t \right)}}$. In Fig.\ref{f4} we plot the difference $\Delta\phi\equiv \delta\phi_{BD}-\delta\phi_{GR}$ as a function of initial value of $r_0$ and redshift, for a set of different $w$.
Our results demonstrate a consistent alignment between a smaller value of $w$ and a more pronounced phase shift, which aligns with theoretical expectations. Notably, when the initial values of $r_0$ are relatively small and $z$ is relatively large, the phase change difference between BD theory and GR become more prominent. Leveraging these distinctions in entanglement phase variations, the QGEM mechanism emerges as a promising observational approach for discerning between gravity theories and cosmological models.

\section{Characteristic spectrum}
Entanglement between particle pairs might have been generated before geodesic motion starts. Therefore, it becomes crucial to determine whether gravity induces entanglement or if other physical processes are involved. To address this challenge, we propose that QGEM during geodesic motion will exhibit a characteristic spectrum as phase changes occur in a series of geodesic lines. By analyzing the entangled patterns formed by different geodesics, we can infer whether the entanglement originates from the gravitational field of nearby particles or from alternative sources. 

Phase itself is not directly observable. We often use entanglement witness $\mathcal{W}$ as an experimental indicator to detect entanglement formation. The definition of entanglement witness is
\beq\label{3}
 \mathcal{W} = \left| {\left\langle {\sigma_{1_x} \otimes \sigma_{2_z}} \right\rangle  + \left\langle {\sigma_{1_y} \otimes \sigma_{2_y}} \right\rangle } \right|.
\eeq
When it's greater than 1, we can infer that there is entanglement between the two particles. 

\begin{figure}\centering
\includegraphics[scale=0.4]{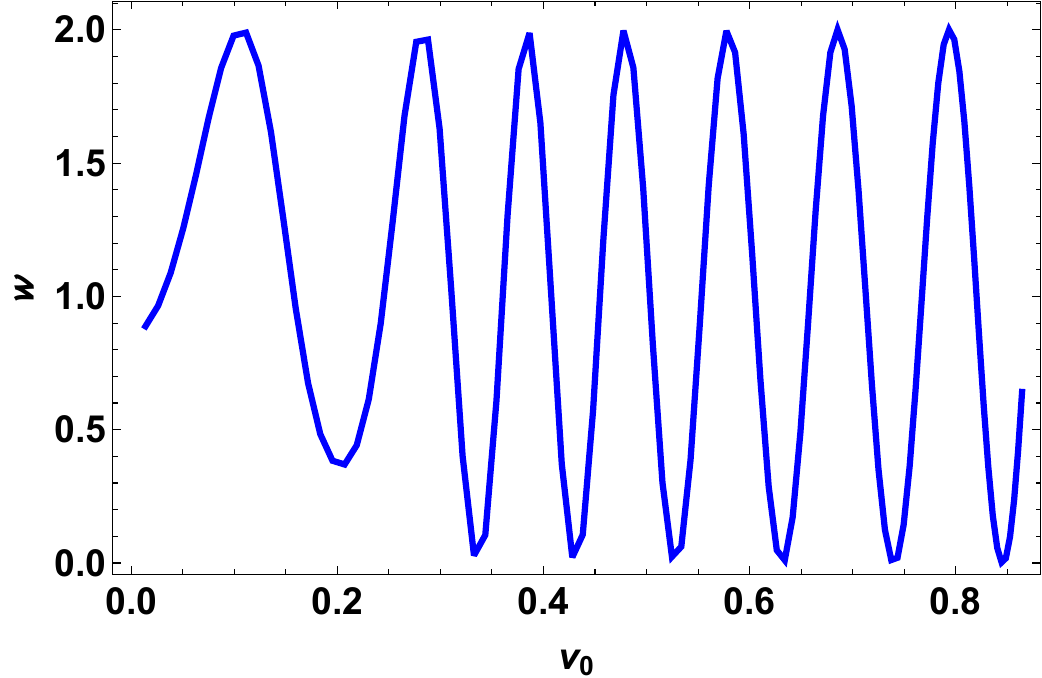}
\caption{Variation of entanglement witness $\mathcal{W}$ as a function of $v_0$ in FLRW universe. Other initial parameters are set to the same as S1 in Table \ref{initial}.}\label{wv0}
\end{figure}

Similar to the case of \cite{Zhang:2023llk}, the entanglement witness $\mathcal{W}$  exhibits oscillatory variations with different initial motion parameters. Again, let's take two series of geodesics as mentioned in Fig. \ref{f7} for exampled. In Fig. \ref{wv0}, we have converted different initial parameters $r_0$ into corresponding initial velocity $u^{\mu}(0)$. While the initial values of four velocity, $u^{\mu}(0)$, can be further transformed to the orthogonal frame form as:
\beq\label{36}
u^{\mu}(\tau=0) = \gamma {e_{(0)}}^{\mu} - \gamma v_0 {e_{(1)}}^{\mu},
\eeq
where $\gamma  = \frac{1}{{\sqrt {1 - {v_0^2}} }}$ and $v_0$ is the initial velocity of the particle with which we replace $r_0$. It shows $v_0$ increases with $r_0$ increases and  $\mathcal{W}$  oscillates faster in the larger area of $v_0$. 

 $\mathcal{W}$ oscillates very unevenly as a function of $r_0$. Only when $r_0$ falls in some certain intervals, the entanglement phase that gives a positive answer to whether to be entangled can be reached. This forces us to limit the particle source to a reasonable spacetime region in order to ensure that entangled particles can be observed when using the entanglement witness \eqref{3}.

\begin{figure}\centering
\includegraphics[scale=0.4]{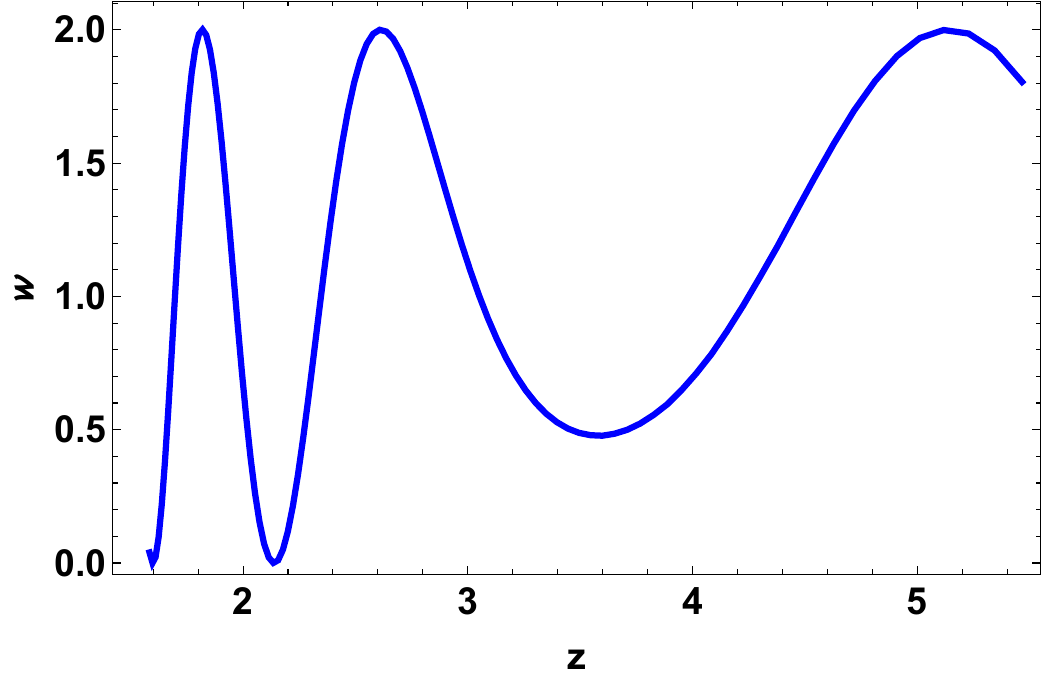}
\caption{Variation of entanglement witness $\mathcal{W}$ as a function of $z$ in FLRW universe. Other initial parameters are set to the same as S1 in Table\ref{initial}}\label{wz}
\end{figure}

Fig. \ref{wz} shows the entanglement witness as a function of the redshift $z$. From where we find that due to the convexity of the phase transition curve S1 in Fig. \ref{f2b}, the entanglement witness curve has a valley in the middle of z that is insufficient to confirm entanglement. At both ends of z, $\mathcal{W}$ oscillates fast enough to reach the threshold, 1, for confirming entanglement.

Compared to the initial conditions, particle's energy is an important observable. In observer's frame, each particle emitted at  different $r_0$, $t_0$ will have a specific geodesic, and every instantaneous observer on the geodesic will measure a corresponding kinetic energy.

\begin{figure}\centering
\includegraphics[scale=0.4]{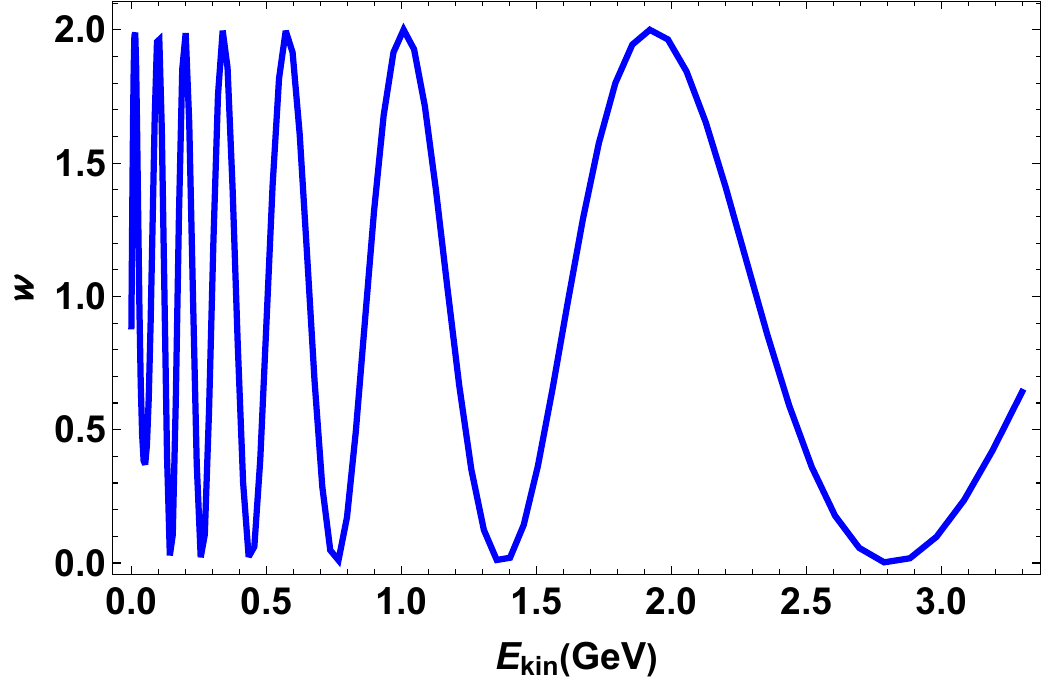}
\caption{Variation of entanglement witness $\mathcal{W}$ as a function of $E_{kin}$ with $t_0$ fixed at $z=3.94$ in FLRW universe. The other kinematic initial conditions are consistent with that in Fig. \ref{wv0}, and each ${v_0}$ corresponds to each ${E_{kin}}$ of this figure.}\label{rew}
\end{figure}

\begin{figure}\centering
	\includegraphics[scale=0.4]{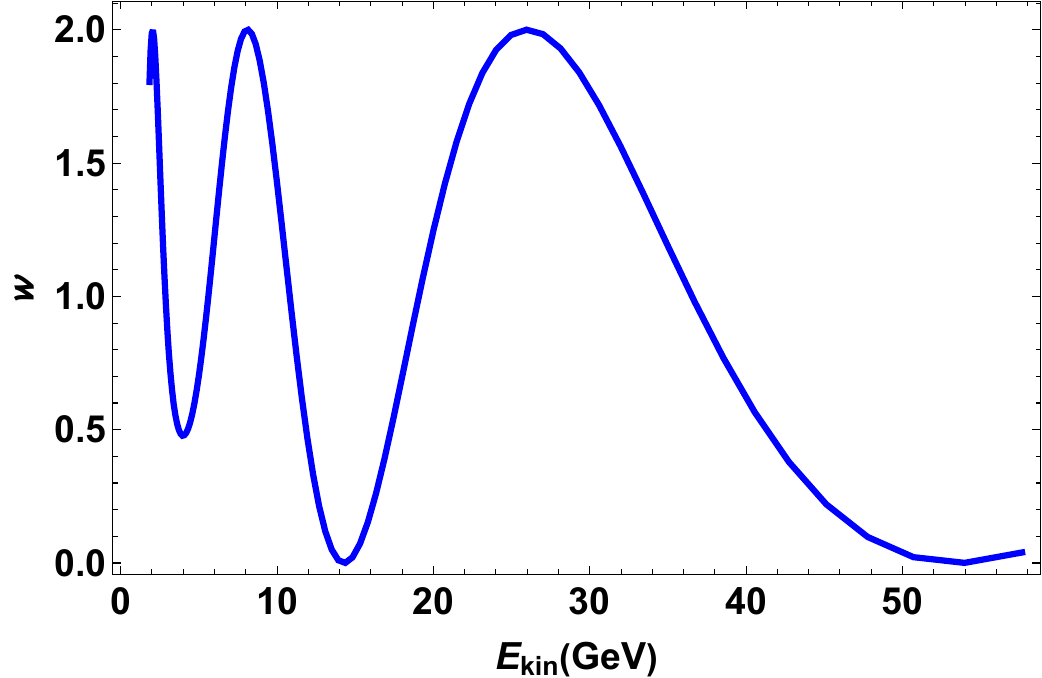}
	\caption{Variation of entanglement witness $\mathcal{W}$ as a function of $E_{kin}$ with fixed ${r_0}$ at $4.5 \times {10^{17}}$ in FLRW universe. The kinematic initial conditions are consistent with that in Fig. \ref{wz}, and each $z$ corresponds to each abscissa ${E_{kin}}$ of this figure.}\label{tew}
\end{figure}

We use the kinetic energy of the particles measured by observer on Earth to label each particle. The observed kinetic energy for particles emitted from different initial spacetime location is calculated as follows:
\bqn\label{6}
{E_{kin}} = {z_a}{p^a} - m,
\eqn
where ${z_a}$ is the four-velocity of an observer on Earth, ${p^a}$ is the four-velocity of a particle upon reaching Earth.
The variation of $\mathcal{W}$ with respect to $E_{kin}$ is plotted in Fig. \ref{rew} and Fig. \ref{tew}. From these two figures, we observe that for both cases, $\mathcal{W}$ oscillates faster in the low kinetic energy ${E_{kin}}$ region than in the high kinetic energy region. Observing low-energy particle pairs will be more helpful for us to detect entanglement. 

\section{Conclusions}
In this paper, we generalize the QGEM experiment to be carried out in FLRW cosmological spacetimes so as to verify quantum gravity after in Schwarzschild spacetimes in \cite{Zhang:2023llk}. Just like in Schwarzschild spacetime, particle pairs propagating in FLRW cosmological spacetime can become entangled as a result of the gravitational field's influence, thus verifying quantum gravity. The entanglement phase is also influenced by variations in its motion parameters, such as the initial launch point position, initial geodesic deviation and its rate. Moreover, the dependence between entanglement curves and cosmological models can be used to verify different gravity theories or cosmological models.

Since the entangled phase cannot be directly observed, in order to connect it more closely with actual cosmological observations, we drew the entanglement witness-energy diagram. Similar to \cite{Zhang:2023llk}, there is a characteristic spectrum of QGEM. So we could distinguish whether the observed entanglement
arises from the quantum gravity effect of particle pairs or from other process (e.g, the emission stage at the source as suggested by the Hawking radiation of black holes).

In addition to verifying the quantum gravity effect, our solution can also verify the extended equivalence principle in the spacetime background of FLRW cosmology \cite{Giacomini:2020ahk}. It also provides a new perspective for the observation of cosmic rays, enriching the content of entanglement detection in cosmology. We provide a new possible source of entanglement formation in cosmology.

In the future, we would like to further study the impact of different cosmological models on the quantum entanglement induced by the QGEM mechanism. The entanglement induced in this way under different cosmological models is very likely to have significantly different characteristics, which will provide a new way for us to observe and study the universe. Moreover, in addition to inducing entanglement between physical particles, quantum gravity can also induce entanglement between quantum fields, such as scalar fields, vector fields, and spinor fields, thus affecting the overall entangled structure of the universe. This could be explored further in the future.
 \section*{Acknowledgements}
This work was supported by the National Natural Science Foundation of China with the Grants No. 12375049, 11975116, and Key Program of the Natural Science Foundation of Jiangxi Province under Grant No. 20232ACB201008. 


\begin{thebibliography}{99}

\bibitem{feynman}R.~Feynman, 
``The role of gravitation in physics'', 
in Chapel Hill Conference Proceedings, edited by C. M.
DeWitt and D. Rickles (Edition Open Access, 1957) pp.
250–256.

\bibitem{r2}
S.~Bose, A.~Mazumdar, G.~W.~Morley, H.~Ulbricht, M.~Toro\v{s}, M.~Paternostro, A.~Geraci, P.~Barker, M.~S.~Kim and G.~Milburn,
``Spin Entanglement Witness for Quantum Gravity,''
Phys. Rev. Lett. \textbf{119}, no.24, 240401 (2017)
[arXiv:1707.06050 [quant-ph]].
\bibitem{r3}
C.~Marletto and V.~Vedral,
``Gravitationally induced entanglement between two massive particles is sufficient evidence of quantum effects in gravity,''
Phys. Rev. Lett. \textbf{119}, no.24, 240402 (2017)
[arXiv:1707.06036 [quant-ph]].

\bibitem{Christodoulou:2018cmk}
M.~Christodoulou and C.~Rovelli,
``On the possibility of laboratory evidence for quantum superposition of geometries,''
Phys. Lett. B \textbf{792}, 64-68 (2019)
[arXiv:1808.05842 [gr-qc]].

\bibitem{Capolupo:2019peg} 
A.~Capolupo, G.~Lambiase, A.~Quaranta and S.~M.~Giampaolo,
``Probing axion mediated fermion\textendash{}fermion interaction by means of entanglement,''
Phys. Lett. B \textbf{804}, 135407 (2020)
[arXiv:1910.01533 [hep-ph]].

\bibitem{Belenchia:2018szb}
A.~Belenchia, R.~M.~Wald, F.~Giacomini, E.~Castro-Ruiz, \v{C}.~Brukner and M.~Aspelmeyer,
``Quantum Superposition of Massive Objects and the Quantization of Gravity,''
Phys. Rev. D \textbf{98}, no.12, 126009 (2018)
[arXiv:1807.07015 [quant-ph]].
\bibitem{Carney:2018ofe}
D.~Carney, P.~C.~E.~Stamp and J.~M.~Taylor,
``Tabletop experiments for quantum gravity: a user\textquoteright{}s manual,''
Class. Quant. Grav. \textbf{36}, no.3, 034001 (2019)
[arXiv:1807.11494 [quant-ph]].
\bibitem{Marshman:2019sne}
R.~J.~Marshman, A.~Mazumdar and S.~Bose,
``Locality and entanglement in table-top testing of the quantum nature of linearized gravity,''
Phys. Rev. A \textbf{101}, no.5, 052110 (2020)
[arXiv:1907.01568 [quant-ph]].
\bibitem{Westphal:2020okx}
T.~Westphal, H.~Hepach, J.~Pfaff and M.~Aspelmeyer,
``Measurement of gravitational coupling between millimetre-sized masses,''
Nature \textbf{591}, no.7849, 225-228 (2021)
[arXiv:2009.09546 [gr-qc]].
\bibitem{Carlesso:2019cuh}
M.~Carlesso, A.~Bassi, M.~Paternostro and H.~Ulbricht,
``Testing the gravitational field generated by a quantum superposition,''
New J. Phys. \textbf{21}, no.9, 093052 (2019)
[arXiv:1906.04513 [quant-ph]].
\bibitem{Danielson:2021egj}
D.~L.~Danielson, G.~Satishchandran and R.~M.~Wald,
``Gravitationally mediated entanglement: Newtonian field versus gravitons,''
Phys. Rev. D \textbf{105}, no.8, 086001 (2022)
[arXiv:2112.10798 [quant-ph]].
\bibitem{Christodoulou:2022mkf}
M.~Christodoulou, A.~Di Biagio, M.~Aspelmeyer, \v{C}.~Brukner, C.~Rovelli and R.~Howl,
``Locally Mediated Entanglement in Linearized Quantum Gravity,''
Phys. Rev. Lett. \textbf{130}, no.10, 100202 (2023)
[arXiv:2202.03368 [quant-ph]].
\bibitem{Cho:2021gvg}
H.~T.~Cho and B.~L.~Hu,
``Quantum noise of gravitons and stochastic force on geodesic separation,''
Phys. Rev. D \textbf{105}, no.8, 086004 (2022)
[arXiv:2112.08174 [gr-qc]].

\bibitem{Matsumura:2020law}
A.~Matsumura and K.~Yamamoto,
``Gravity-induced entanglement in optomechanical systems,''
Phys. Rev. D \textbf{102}, no.10, 106021 (2020)
[arXiv:2010.05161 [gr-qc]].

\bibitem{Miki:2020hvg}
D.~Miki, A.~Matsumura and K.~Yamamoto,
``Entanglement and decoherence of massive particles due to gravity,''
Phys. Rev. D \textbf{103}, no.2, 026017 (2021)
[arXiv:2010.05159 [gr-qc]].

\bibitem{Howl:2023xtf}
R.~Howl, N.~Cooper and L.~Hackerm\"uller,
``Gravitationally-induced entanglement in cold atoms,''
[arXiv:2304.00734 [quant-ph]].

\bibitem{Yant:2023smr}
J.~Yant and M.~Blencowe,
``Gravitationally induced entanglement in a harmonic trap,''
Phys. Rev. D \textbf{107}, no.10, 106018 (2023)
[arXiv:2302.05463 [quant-ph]].

\bibitem{Feng:2023krm}
S.~Feng, B.~M.~Gu and F.~W.~Shu,
``Detecting Extra Dimension By the Experiment of the Quantum Gravity Induced Entanglement of Masses,''
[arXiv:2307.11391 [gr-qc]].

\bibitem{Schut:2023eux}
M.~Schut, A.~Grinin, A.~Dana, S.~Bose, A.~Geraci and A.~Mazumdar,
``Relaxation of experimental parameters in a Quantum-Gravity Induced Entanglement of Masses Protocol using electromagnetic screening,''
[arXiv:2307.07536 [quant-ph]].

\bibitem{Fragolino:2023agd}
P.~Fragolino, M.~Schut, M.~Toro\v{s}, S.~Bose and A.~Mazumdar,
``Decoherence of a matter-wave interferometer due to dipole-dipole interactions,''
[arXiv:2307.07001 [quant-ph]].

\bibitem{Li:2022yiy}
P.~Li, Y.~Ling and Z.~Yu,
``Generation rate of quantum gravity induced entanglement with multiple massive particles,''
Phys. Rev. D \textbf{107}, no.6, 064054 (2023)
[arXiv:2210.17259 [gr-qc]].

\bibitem{Bose:2022uxe}
S.~Bose, A.~Mazumdar, M.~Schut and M.~Toro\v{s},
``Mechanism for the quantum natured gravitons to entangle masses,''
Phys. Rev. D \textbf{105}, no.10, 106028 (2022)
[arXiv:2201.03583 [gr-qc]].

\bibitem{He:2023hys}
F.~He and B.~Zhang,
``Generation of entanglement between two laser pulses through gravitational interaction,''
Eur. Phys. J. Plus \textbf{138}, no.2, 141 (2023)
[arXiv:2302.06362 [gr-qc]].

\bibitem{Zhang:2023llk}
C.~Zhang and F.~W.~Shu,
``Gravity-induced entanglement between two massive microscopic particles in curved spacetime: I.The Schwarzschild background,''
[arXiv:2308.16526 [gr-qc]].

\bibitem{Martin-Martinez:2012chf}
E.~Martin-Martinez and N.~C.~Menicucci,
``Cosmological quantum entanglement,''
Class. Quant. Grav. \textbf{29}, 224003 (2012)
[arXiv:1204.4918 [gr-qc]].

\bibitem{Brahma:2020zpk}
S.~Brahma, O.~Alaryani and R.~Brandenberger,
``Entanglement entropy of cosmological perturbations,''
Phys. Rev. D \textbf{102}, no.4, 043529 (2020)
[arXiv:2005.09688 [hep-th]].

\bibitem{Kiefer:2008ku}
C.~Kiefer and D.~Polarski,
``Why do cosmological perturbations look classical to us?,''
Adv. Sci. Lett. \textbf{2}, 164-173 (2009)
[arXiv:0810.0087 [astro-ph]].

\bibitem{Martin:2007bw}
J.~Martin,
``Inflationary perturbations: The Cosmological Schwinger effect,''
Lect. Notes Phys. \textbf{738}, 193-241 (2008)
[arXiv:0704.3540 [hep-th]].

\bibitem{Martin:2015qta}
J.~Martin and V.~Vennin,
``Quantum Discord of Cosmic Inflation: Can we Show that CMB Anisotropies are of Quantum-Mechanical Origin?,''
Phys. Rev. D \textbf{93}, no.2, 023505 (2016)
[arXiv:1510.04038 [astro-ph.CO]].

\bibitem{Martin:2016tbd}
J.~Martin and V.~Vennin,
``Bell inequalities for continuous-variable systems in generic squeezed states,''
Phys. Rev. A \textbf{93}, no.6, 062117 (2016)
[arXiv:1605.02944 [quant-ph]].

\bibitem{Lello:2013bva}
L.~Lello, D.~Boyanovsky and R.~Holman,
``Entanglement entropy in particle decay,''
JHEP \textbf{11}, 116 (2013)
[arXiv:1304.6110 [hep-th]].

\bibitem{Valentini:2006yj}
A.~Valentini,
``Astrophysical and cosmological tests of quantum theory,''
J. Phys. A \textbf{40}, 3285-3303 (2007)
[arXiv:hep-th/0610032 [hep-th]].

\bibitem{Chen:2017cgw}
J.~W.~Chen, S.~H.~Dai, D.~Maity, S.~Sun and Y.~L.~Zhang,
``Towards Searching for Entangled Photons in the CMB Sky,''
Phys. Rev. D \textbf{99}, no.2, 023507 (2019)
[arXiv:1701.03437 [quant-ph]].

\bibitem{r7} R. D. Inverno, Introducing Einstein's Relativity (Clarendon Press, Oxford, 1992).

\bibitem{Aghanim:2018eyx}
N.~Aghanim \textit{et al.} [Planck],
``Planck 2018 results. VI. Cosmological parameters,''
Astron. Astrophys. \textbf{641}, A6 (2020)
[erratum: Astron. Astrophys. \textbf{652}, C4 (2021)]
[arXiv:1807.06209 [astro-ph.CO]].

\bibitem{Gaztanaga:2000vw}
E.~Gaztanaga and J.~A.~Lobo,
Astrophys. J. \textbf{548}, 47-59 (2001)
doi:10.1086/318684
[arXiv:astro-ph/0003129 [astro-ph]].

\bibitem{Clifton:2011jh}
T.~Clifton, P.~G.~Ferreira, A.~Padilla and C.~Skordis,
Phys. Rept. \textbf{513}, 1-189 (2012)
doi:10.1016/j.physrep.2012.01.001
[arXiv:1106.2476 [astro-ph.CO]].

\bibitem{Giacomini:2020ahk}
F.~Giacomini and \v{C}.~Brukner,
``Einstein's Equivalence principle for superpositions of gravitational fields and quantum reference frames,''
[arXiv:2012.13754 [quant-ph]].


\end{thebibliography}
\end{document}